\documentclass{article}%
\usepackage{makeidx}
\usepackage{amsmath}
\usepackage{amsfonts}
\usepackage{amssymb}
\usepackage{graphicx}%
\begin{document}

\title{Comments to the article "Parametric fitting of data obtained from detectors
with finite resolution and limited acceptance" by Gagunashvili \cite{gagu}}
\author{G. Bohm\thanks{Deutsches Elektronen-Synchrotron, DESY, D-15738 Zeuthen,
Germany, email:bohm@ifh.de} and G. Zech\thanks{Universit\"at Siegen, D-57068
Siegen, Germany, email: zech@physik.uni-siegen.de}}
\maketitle

\begin{abstract}
The publication \cite{gagu} suffers from several caveats: i) The method is
based upon the false assumption that the median of $\chi^{2}$ distributed
random variables is $\chi^{2}$ distributed. ii) The information contained in
the data is not fully used, iii) It is not clear how the uncertainties
associated to the fitted parameters can be evaluated. A correct solution of
the problem is presented and results of Ref. \cite{gagu} are compared to
results obtained using the approach described in our textbook \cite{bohm}.
Finally, we correct false statements in \cite{gagu} about a section in our book.

\end{abstract}




\section{Introduction}

The determination of parameters of a theory from experimental data that are
distorted by experimental effects is a relatively simple problem. This is why
the solution of this problem had not been published in a referenced scientific
journal so far, but last year Gagunashvili has submitted the cited paper.

The problem is solved in the following way: The experimental data are compared
to a Monte Carlo simulation in form of histograms of the observed variable
$x^{\prime}$ which due to the finite experimental resolution differs from the
true variable $x$. A $\chi^{2}$ expression is formed that measures the
statistical difference between the two histograms. The simulated histogram
depends on the parameter $\theta$ of interest. ($\theta$ may also represent a
set of parameters.) In a least square (LS) fit the parameters are estimated.
The effect of changing the parameters is implemented by changing the weights
of the simulated events, $w(x)=f(x|\theta)/f(x|\theta_{0})$ where $x$ is the
undistorted variable, $\theta$ the parameter of interest of the p.d.f.
$f(x|\theta)$ and $\theta_{0}$ its value used in the simulation. This is
explained in more detail in Ref. \cite{bohm}.

The purpose of this comment is twofold: i) We want to point out some caveats
of the treatment in Ref. \cite{gagu} which will lead in some applications to
biased results and wrong error assignments. A correct solution is presented.
ii) We want to correct several false statements in this article in respect to
our textbook \cite{bohm}. In a numerical example we compare our results to
those published in Ref. \cite{gagu}.

\section{The $\chi^{2}$ approximation - multinomial versus Poisson
distribution}

To illustrate the problem, we look at a simple example: In an experiment the
slope of a linear distribution is to be measured. A certain amount of data has
been accumulated and distributed into histogram bins. The number of events in
each bin follows a Poisson distribution the mean of which depends on the flux
and the slope of the distribution. Usually the flux is not known and thus we
have two unknown parameters, the slope and the normalization of our prediction
to the data. The likelihood principle tells us that it does not matter whether
the experiment has been stopped after a certain running time or after a
certain amount of data had been collected. This follows from the likelihood
principle and has been shown explicitly in Ref. \cite{cousins} for the maximum
likelihood estimate (MLE). Asymptotically, in the Gaussian approximation which
we use below, the least square (LS) expression coincides with half the
negative log-likelihood function up to irrelevant terms and thus also the LS
estimate is independent of the stopping condition. In Ref. \cite{gagu} is
assumed that the number of events is fixed and a multinomial approach is
applied. This condition is rarely realized and does not require the
multinomial treatment but the latter is correct, too.

For a Poisson distributed number of events $d_{i}$ in a histogram bin $i$ and
predictions $t_{i}$ which depend on parameters of interest, the distribution
$W(d_{1},..,d_{B})$ in the histogram with $B$ bins is described by%

\begin{equation}
W=\prod\limits_{i=1}^{B}P(d_{i}|t_{i}) \label{poisson}%
\end{equation}
where $P(m|\lambda)$ is the Poisson distribution of $m$ with mean $\lambda$.
It can also be written as%

\begin{equation}
W=P(d|t)M(d_{1},..,d_{B}|p_{1},..,p_{B},d) \label{poissonmulti}%
\end{equation}
with $d=\Sigma_{i=1}^{B}d_{i}$, $t=\Sigma_{i}^{B}t_{i}$, $p_{i}=t_{i}/t$ and
$M$ the multinomial distribution for $d$ events distributed into $B$ bins with
probabilities $p_{1},..,p_{B}$. Relations (\ref{poisson}) and
(\ref{poissonmulti}) describe the same distribution. We are free to use either
form (see also Ref. \cite{kendall}), but (\ref{poisson}) is to be preferred in
our problem because the formulas become much simpler than with
(\ref{poissonmulti}). In \cite{gagu} $\chi^{2}$ is evaluated based upon
multinomial statistics, while in \cite{bohm} Poisson statistics has been
applied. Both ways are correct, but the multinomial approach is involved
because correlations have to be handled explicitly.

If only relative predictions are available, we have to normalize the
predictions. From the factorization of (\ref{poissonmulti})\ we find the
maximum likelihood estimate $\hat{t}=d$ which is introduced into
(\ref{poisson}), $\Sigma t_{i}=\Sigma d_{i}=d$.

To compare $B$ Poisson distributed numbers $d_{i}$ to a prediction $t_{i}$, we
form the statistic $\chi^{2}=\Sigma_{i=1}^{B}[(d_{i}-t_{i})^{2}/t_{i}]$ which
follows a $\chi^{2}$ distribution with $B$ degrees of freedom (NDF) in the
approximation where for each bin $i$ the Poisson distribution of $d_{i}$ with
mean $t_{i}$ can be approximated by a normal distribution with mean and
variance $t_{i}$. When the predictions $t_{i}$ are normalized, i.e. $\Sigma
d_{i}=\Sigma t_{i}$, we loose one degree of freedom ($1$ parameter estimated)
and have $NDF=B-1$. If $P$ parameters of interest have to be estimated, we
have $NDF=B-P-1$

If we have a Monte Carlo prediction consisting of $K_{i}$ events in bin $i$
with weights $w_{ik}$, we have $t_{i}=c\Sigma_{k}w_{ik}$ where $c$ is an
overall normalization constant and we get:%

\begin{equation}
\chi^{2}=%
{\displaystyle\sum\limits_{i=1}^{B}}
\chi_{i}^{2}=%
{\displaystyle\sum\limits_{i=1}^{B}}
\frac{(d_{i}-c\Sigma_{k}w_{ik})^{2}}{c\Sigma_{k}w_{ik}}\;. \label{chi1}%
\end{equation}

When we estimate the parameters hidden in the weights, also $c$ is a free
parameter in the fit and as above $NDF=B-P-1$. In (\ref{chi1}) the relative
statistical error of $t_{i}$ is assumed to be small compared to $d_{i}^{-1/2}%
$. This covers probably more than 90 \% of all cases in particle physics applications.

If the statistical error of the simulation cannot be neglected, we consider
the asymptotic case where not only the distributions of $d_{i}$ but also that
of $\Sigma_{k}w_{ik}$ can be approximated by normal distributions. We form

\begin{equation}
\chi^{2}=%
{\displaystyle\sum\limits_{i=1}^{B}}
\frac{(d_{i}-c\Sigma_{k}w_{ik})^{2}}{\sigma_{i}^{2}}\;.
\end{equation}

To obtain a $\chi^{2}$ expression, the quantity $\sigma_{i}^{2}$ per
definition has to be the variance of $d_{i}-c\Sigma_{k}w_{ik}$ under the
assumption that the prediction is correct. A reasonable approximation is
obtained from error propagation, $\hat{\sigma}_{i}^{2}=d_{i}+c^{2}\Sigma
_{k}w_{ik}^{2}$. This approximation is adequate in most of the remaining
cases. A correct treatment includes $\sigma$ in the fit. A maximum likelihood
estimate (see Appendix 13, subsection 13.8.1 and relation (13.32) of Ref.
\cite{bohm} in different notation and Ref. \cite{bohm1}) is:%

\begin{equation}
\hat{\sigma}_{i}^{2}=c(d_{i}\frac{\Sigma_{k}w_{ik}^{2}}{\Sigma_{k}w_{ik}%
}+\Sigma_{k}w_{ik})\;. \label{sigmaml}%
\end{equation}

To derive (\ref{sigmaml}) the expected values $\mathbb{E}(w_{ik})$ and
$\mathbb{E}(w_{ik}^{2})$ have been replaced by the empirical values
$\Sigma_{k}w_{ik}/K_{i}$ and $\Sigma_{k}w_{ik}^{2}/K_{i}$. For convenience,
the derivation of (\ref{sigmaml}) is given in the Appendix.

The normalization $c$ is a free parameter in the fit. Asymptotically, the
fitted normalization reproduces the number of observed events.

\section{Caveats, restrictions and difficulties of the approach described in
\cite{gagu}.}

\subsection{The choice of the median}

The difficulties in the approach in \cite{gagu} are related to the application
of the multinomial distribution where the correlations have to be explicitly
handled. There, $\chi_{i}^{2}$ is evaluated excluding bin $i$ for all $B$
values of $i$ and then the median is computed without any explanation and
discussion of its properties.

The ad hoc solution to take the median instead of the mean of many incomplete
evaluations is used because the median is a robust estimator - an indication
that something is problematic with this approximation, but both choices are
wrong: \emph{Neither the mean nor the median of the }$\chi_{NDF}^{2}$\emph{
distributed values is described by a }$\chi_{NDF}^{2}$\emph{ distribution as
long as the values are not identical}. For a typical example taken from
\cite{gagu1} and using \ Relation (34) of \cite{gagu1}, we find that the mean
value of the medians of samples of partially correlated random variables
following a $\chi_{B-2}^{2}$ distribution is typically by half a unit off the
nominal value of $NDF=B-2$. This bias does not disappear with increasing
statistics. As a consequence the evaluation of parameter uncertainties from
the curvature of the $\chi^{2}$ curve near its maximum is doubtful and
p-values derived from a $\chi^{2}$ test are wrong. The fact that in specific
examples the point estimate is sensible does not validate the proposed procedure.

\subsection{The loss of a constraint}

Comparing a prediction of a histogram to a histogram of observed events where
the prediction is to be normalized to the data, we have $NDF=B-1-P$ degrees of
freedom, where $P$ is the number of additional parameters of interest. In
\cite{gagu} due to the difficulties created by the correlations, one
additional degree of freedom is given away, $NDF=B-2-P$. As a consequence,
precision is lost and, for example, an asymmetry in a two-bin histogram cannot
be determined.

\subsection{The weight restriction}

The weights in \cite{gagu} are restricted to functions of the histogrammed
variable. This condition is violated in unfolding problems. The weight
restriction seems not to be necessary for unnormalized weights but this point
should be clarified officially by the author.

\subsection{The error treatment}

\emph{In \cite{gagu} it is not explained how the parameter errors are obtained
in the fitting procedure.} A parameter fit without error assignment is
useless. The reader will probably assume that the errors are obtained in the
standard way from the variation of $\chi^{2}$ as a function of the parameter
(at some point MINUIT errors are quoted). However, the standard error
estimation can be wrong independent of the fact that in the evaluation of the
$\chi^{2}$ statistic the errors are correctly implemented as a function of the
parameter values. The reason is explained in our book: It is related to the
fact that due to the weighting the denominators of the $\chi^{2}$ expression
may have a sizable dependence of the parameters. The effect is small in most
applications but can be large, if histograms with a large number of bins and
large smearing are fitted. Therefore, in situations where the uncertainty of
the simulated numbers cannot be neglected, the validity of the error
assignment has to be checked, for instance by changing the amount of Monte
Carlo events.

In addition to the problems related to weighting, the error handling in least
square fits where crude approximations of the $\chi^{2}$ statistic are used
should have been discussed. The effects are especially important with low
event numbers.

\subsection{Problems with small event numbers}

In many experiments the number of events is so small that the application of a
least square fit is problematic. In \cite{gagu} \ no solution for this
situation is presented. However in this case enough Monte Carlo events can be
generated such that their statistical error is negligible and a Poisson
\ likelihood fit can be performed as explained in Ref. \cite{bohm,bohm1}.

\subsection{Technical difficulties}

In the approach of \cite{gagu} a parameter fit with a histogram of $B$ bins
requires for each change of the parameter during the minimum search $B$
additional fits of an auxiliary parameter. For a two-dimensional histogram
with $20\times20$ bins and $1000$ minimum searching steps in the fit this
means that $400,000$ auxiliary parameters have to be estimated in LS fits.

\section{Inflicting statements about the content of our book}

In Section 1 of Ref. \cite{gagu} figures the following paragraph:

\textquotedblleft In [2], a re-weighting procedure for fitting a Monte Carlo
reconstructed distribution to the reconstructed data was proposed. The
procedure is presented rather sketchily, and cannot be repeated even for the
example that was used in [2] for illustration. There is not a clear
explanation of how the parameters and the errors in them were calculated. The
authors of [2] stated without proof, that the statistic used for the fitting
of the parameters had a $\chi^{2}$ distribution but did not define the number
of degrees of freedom. This makes it impossible to use this statistic for
choosing the best model from a set of alternative models.\textquotedblright

The reference was our book Ref. \cite{bohm}.

These statements are false:

- \emph{Two methods are proposed. The likelihood ratio solution is not
mentioned} in Ref. \cite{gagu}. \emph{There are two examples}. May be, the
explanations were a bit short but certainly understandable in the context of
preceding chapters of our textbook. We would have been glad to furnish further
explanations to Gagunashvili. Our method is considerably simpler than that
proposed in Ref. \cite{gagu}, certainly not less precise and we are convinced
that it is easier to understand than that presented in Ref. \cite{gagu}.

- Contrary to the statement of Ref. \cite{gagu}, in the two examples\emph{ no
parameter estimates and errors were quoted}. The section the author of Ref.
\cite{gagu} refers to is Section 6.5.9, "Comparison of Observations with a
Monte Carlo Simulation". It is part of a chapter on parameter inference in
which it is \emph{explained in detail how parameters and their errors are
estimated in least square and likelihood fits} (see Sections 6.5.3 and 6.5.5,
subsection $\chi^{2}$ approximation). A subsequent chapter on interval
estimation discusses error assignments even in more detail.

- The author apparently refers to formula (6.17) in different notation of
Section 6.5.9 of our textbook:%

\begin{equation}
\chi^{2}=\sum\limits_{i=1}^{B}\frac{(d_{i}-cm_{i})^{2}}{cm_{i}} \label{chi}%
\end{equation}
where $d_{i}$ was as the number of experimental events in bin $i$, $cm_{i}$
the Monte Carlo prediction with $c$ a Monte Carlo normalization constant and
$m_{i}$ the corresponding sum of weights in bin $i$, $B$ was the number of
bins. It was stated that the formula is valid if the statistical uncertainty
of $m_{i}$ is negligible. The formula was derived and explained in the section
"$\chi^{2}$ approximation" where $\chi^{2}$ for a histogram with Poisson
distributed numbers was discussed. \emph{The NDF are irrelevant for parameter
estimation}. (The NDF has to be known in goodness-of-fit tests which is a
different subject and which is treated in a subsequent chapter of our book.
Independent of this fact,\emph{ the number of degrees of freedom (NDF) for
}$\chi^{2}$\emph{ statistics depending on fitted parameters were defined in
Chapter 3} and therefore are known to the reader. The relation needed to apply
a goodness-of-fit test for weighted histograms (13.32) is given in the
Appendix 13 where also the NDF are defined. The relation needed for parameter
estimation in the case that the uncertainty of the simulation has to be taken
into account is given in the subsection 13.8.4 but the statements in
\cite{gagu} referred to the main text.)

\section{Example}

To perform a quantitative comparison of our approach to that of Ref.
\cite{gagu}, we have applied our method to the first example given in Ref.
\cite{gagu} which was also used in our book to illustrate our method. The
slope of a linear distribution is to be adjusted. The p.d.f. is $f(x|\alpha
)=(1+\alpha x)/(1+\alpha/2)$, with $x\in\lbrack0,1]$ and $\alpha>-1$. For the
\textquotedblleft experimental\textquotedblright\ events the slope parameter
was $\alpha=1$ and for the Monte Carlo events $\alpha_{m}=0$. Events were
generated in the interval $[0,1]$. The variable $x$ was smeared with a
Gaussian resolution of $\sigma=0.3$. The smeared distribution was subdivided
into $5$ or $\ 20$ bins of equal width in the range between $-0.3$ and $1.3$.
The number of experimental and simulated events was $500$, $5000$ and $50000$.
Each case was simulated $10,000$ times. Some choices required to take the
error of the simulation into account. We applied formula (13.32) of Ref.
\cite{bohm} which corresponds to Eq. (\ref{sigmaml}). The slope parameter is
hidden in the weights $w_{k}$. In the least square fit we included only bins
with more than $5$ events. Not all combinations quoted in Ref. \cite{gagu}
were repeated.

We compare our results to those of Ref. \cite{gagu} in Table 1. The results of
Ref. \cite{gagu} are quoted in parenthesis. In the lines denoted by
\textquotedblleft+\textquotedblright\ and \textquotedblleft$-$%
\textquotedblright\ some \ kind of estimates of the positive and negative
errors as defined in Ref. \cite{gagu} are given. In addition to the mean of
the slope parameter which has the nominal value one, we quote the root mean
square deviation (r.m.s.) of the distribution of the fitted slope parameter.
For high statistics our and the results of Ref. \cite{gagu} agree, for small
event numbers we observe mostly a smaller bias and obtain smaller errors. The
results for the specific case with $5$ bins, $500$ observed and $500$
simulated events are unstable because arbitrarily large parameter values occur
if \ the number of experiments is continuously increased. Also other cases
with small observed or simulated event numbers may suffer from some rare cases
where large slopes are found. Therefore we are not sure that all differences
that we observe are really significant. Anyway, least square fits based on a
$\chi^{2}$ approximation are problematic with such low event numbers. For
physics applications only the last column of the table with $20$ bins and
$50000$ simulated events is relevant.

We have also applied a maximum likelihood fit. The results were similar to
those from the least square fit.

\begin{table}[ptb]
\caption{Results from fitting the linear slope parameter $\alpha$ (with
nominal value unity, see text) with our method compared with the results from
\cite{gagu} (in parenthesis) for various sample sizes and bin numbers. Besides
mean and root mean square (rms) values some positive and negative $+,\;-$
error estimates as defined in \cite{gagu} are given.}%
\centering
\begin{tabular}
[c]{|l|l|l|l|l|l|}\hline
\# data &  & 5 bins & 5 bins & 20 bins & 20 bins\\\hline
& \# MC & 500 & 50000 & 500 & 50000\\\cline{2-6}
& mean & 1.25 (1.29) & 1.11 (1.13) & 1.10 (1.17) & 1.00 (1.07)\\
500 & + & 3.37 (3.13) & 0.91 (0.81) & 1.65 (1.84) & 0.72 (0.79)\\
& $-$ & 0.60 (0.66) & 0.46 (0.54) & 0.59 (0.61) & 0.43 (0.45)\\
& rms & 1.15 & 0.65 & 1.14 & 0.55\\\hline
& mean & 1.12 (1.12) & 1.01 (1.01) & 1.11 (1.11) & 1.01 (1.00)\\
5000 & + & 0.87 (0.93) & 0.21 (0.23) & 0.89 (0.91) & 0.20 (0.20)\\
& $-$ & 0.48 (0.52) & 0.16 (0.17) & 0.47 (0.47) & 0.15 (0.16)\\
& rms & 0.66 & 0.19 & 0.64 & 0.18\\\hline
& mean & 1.10 (1.10) & 1.00 (1.00) & 1.11 (1.10) & 1.00 (1.00)\\
50000 & + & 0.86 (0.87) & 0.08 (0.09) & 0.77 (0.83) & 0.08 (0.08)\\
& $-$ & 0.46 (0.49) & 0.07 (0.08) & 0.46 (0.45) & 0.07 (0.07)\\
& rms & 0.63 & 0.08 & 0.62 & 0.07\\\hline
\end{tabular}
\end{table}

\section{Conclusions}

The comparison of histograms of statistical data with a Monte Carlo prediction
should be treated in the framework of Poisson statistics. The correlation of
the event numbers in the different bins can be taken into account by the
normalization of the data to the prediction in the fit. This approach is
considerably simpler than the method of Ref. \cite{gagu} which starts from a
multinomial distribution.

The treatment of Ref. \cite{gagu} is based on the false assumption that the
median of a sample of $\chi^{2}$ distributed random variables also follows a
$\chi^{2}$ distribution with the same number of degrees of freedom. As a
consequence, error estimates based on the corresponding statistic are wrong.
Furthermore the result does not exploit the full information of the data in
that it gives away one degree of freedom. Small event number cannot be handled
and there is no error treatment. In the quantitative comparison of the point
estimates of the two approaches in a special example published in Ref.
\cite{gagu}, the results are found to be similar but they are slightly more
precise in our method.

We reject the false assertions made in Ref. \cite{gagu} with respect to our book.

\section{Appendix: Proof of the relation (\ref{sigmaml})}

We prove relation (\ref{sigmaml}) for a single bin and drop the bin
index. We consider the quantity $d-cK\bar{w}$, where $d$ and $K$ are Poisson
distributed, $\bar{w}$ is the mean value of $K$ weights and $c$ is a
normalization constant common to all bins. In the limit where $d,K$ approach
infinity, the statistic
\begin{equation}
\chi^{2}=\frac{(d-cK\bar{w})^{2}}{c(d\mathbb{E}(w^{2})/\mathbb{E}%
(w)+\mathbb{E}(K)\mathbb{E}(w))}\label{app1}%
\end{equation}
is $\chi^{2}$ distributed with one degree of freedom. Equivalently,
$\sqrt{\chi^{2}}$ is normally distributed with variance equal to one.

\emph{Proof:}

We set%
\begin{equation}
t=K\bar{w}\;\;,\;\;\tilde{t}=t\mathbb{E}(w)/\mathbb{E}(w^{2}),\;\;\tilde
{c}=c\mathbb{E}(w^{2})/\mathbb{E}(w)\;. \label{app2}%
\end{equation}
Relation (\ref{app1}) now reads%
\[
\chi^{2}=\frac{(d-\tilde{c}\tilde{t})^{2}}{c(d\mathbb{E}(w^{2})/\mathbb{E}%
(w)+\mathbb{E}(t))}\;.
\]
According to the central limit theorem, $d,t$ and $\tilde{t}$ are
asymptotically normally distributed. Furthermore we have $Var(\tilde
{t})=\mathbb{E}(\tilde{t})$ which is typical for the Poisson distribution. As
asymptotically the Poisson distribution approaches the normal distribution, we
are allowed to use in the following the Poisson approximation of the
distribution of $\tilde{t}$ instead of the normal distribution. (Without proof
we claim that the Poisson approximation is closer to the true distribution
than the normal approximation. Examples can be found in Ref. \cite{bohm}.)

The variance $\sigma^{2}$ of $d-\tilde{c}\tilde{t}$ is%

\[
\sigma^{2}=\hbox{var}(d)+\tilde{c}^{2}\hbox{var}(\tilde{t})\;.
\]

Under the hypothesis that our description is correct, $d$ and $\tilde{t}$
should follow related Poisson distributions with mean $\lambda$ and
$\lambda/\tilde{c}$, respectively and
\[
\sigma^{2}=\lambda(1+\tilde{c})\;.
\]

We fix $\lambda$ to its maximum likelihood estimate from the log-likelihood
function derived from the product of the two Poisson likelihoods:%

\[
\ln L=d\ln\lambda-\lambda+\tilde{t}\ln\frac{\lambda}{\tilde{c}}-\frac{\lambda
}{\tilde{c}}\;.
\]

Determining $\hat{\lambda}$ as usual from the root of $\partial\ln
L/\partial\lambda$ and re-substituting $c$ using (\ref{app2}) we get%

\begin{subequations}
\begin{equation}
\hat{\lambda}=\frac{\tilde{c}}{1+\tilde{c}}(d+\tilde{t}) \label{lambda}%
\end{equation}
and with%
\end{subequations}
\[
\sigma^{2}=\tilde{c}(d+\tilde{t})=c(d\mathbb{E}(w^{2})/\mathbb{E}%
(w)+\mathbb{E}(t))
\]
the assertion (\ref{app1}).

We replace the expected values by their empirical estimates and obtain
(\ref{sigmaml})$.$

\emph{Remark 1:} A different and more complicated estimate of $\sigma$ is
obtained if we use the normal approximations for the distributions of $d$ and
$t$, but in the asymptotic limit the different approximations coincide.
Therefore our result does not depend on the use of the Poisson approximation
for the distribution of the weighted sum.

\emph{Remark 2:} If $t(\theta)$ depends on a parameter $\theta$ that has to be
estimated, the parameter $\lambda$ and in principle also $\mathbb{E}%
(w),\mathbb{E}(w^{2})$ are nuisance parameters. Estimating them out is correct
for the point estimate of $\theta$. For the interval estimate the three
nuisance estimates depend on $\theta$. Eliminating them with Rel.
(\ref{lambda}) and taking the empirical means corresponds to a profile
likelihood treatment of the error estimate of $\theta$.


\begin{thebibliography}{9}                                                                                                %


\bibitem {gagu}N.D. Gagunashvili, \emph{Parametric fitting of data obtained
from detectors with finite resolution and limited acceptance}, Nucl. Instr.
and Meth. A635 (2011) 88.

\bibitem {bohm}G. Bohm and G. Zech, \emph{Introduction to Statistics and Data
Analysis for Physicists}, Verlag Deutsches Elektronen-Synchrotron, Hamburg
(2010), http://www-library.desy.de/elbooks.html.

\bibitem {cousins}S. Baker and R. D. Cousins, \emph{Clarification of the Use
of Chi-Square and the Likelihood Function in Fits to Histograms}, Nucl. Instr.
and Meth. 221 (1984) 437.

\bibitem {kendall}M. G. Kendall and A. Stuart, \emph{The advanced theory of
statistics}, Charles Griffin, London, 3rd edition, Vol 2, p436ff (1969).

\bibitem {bohm1}G. Bohm and G. Zech, \emph{Comparing statistical data to Monte
Carlo simulation with weighted events}, accepted for publication in Nucl.
Instr. and Meth. A, (2012).

\bibitem {gagu1}N.D. Gagunashvili, \emph{Goodness of fit tests for weighted
histograms}, Nucl. Instr. and Meth. A596 (2008) 439.
\end{thebibliography}
\end{document}